\documentclass[showpacs,preprintnumbers,amsmath,amssymb]{revtex4}
\usepackage{amsmath}
\usepackage{graphicx}
\usepackage{subfig}
\usepackage{hyperref}
\usepackage{amssymb}
\usepackage[english]{babel}
\usepackage{epsfig}
\usepackage{wasysym}

\begin{document}
\title{Generalization of Regular Black Holes in General Relativity to $f(R)$ Gravity\\}
\author{Manuel E. Rodrigues$^{(a,b)}$}\email{esialg@gmail.com}
\author{J\'{u}lio C. Fabris$^{(e,f)}$}\email{fabris@pq.cnpq.br}
\author{Ednaldo L. B. Junior$^{(b,c)}$}\email{ednaldobarrosjr@gmail.com}
\author{Glauber T. Marques$^{(d)}$}\email{gtadaiesky@hotmail.com}
\affiliation{$^{(a)}$ Faculdade de Ci\^{e}ncias Exatas e Tecnologia, 
Universidade Federal do Par\'{a}\\
Campus Universit\'{a}rio de Abaetetuba, CEP 68440-000, Abaetetuba, Par\'{a}, 
Brazil} 

\affiliation{$^{(b)}$ Faculdade de F\'{\i}sica, PPGF, Universidade Federal do 
 Par\'{a}, 66075-110, Bel\'{e}m, Par\'{a}, Brazil}

\affiliation{$^{(c)}$ Faculdade de Engenharia da Computa\c{c}\~{a}o, 
Universidade Federal do Par\'{a}, Campus Universit\'ario de Tucuru\'{\i}, CEP: 
68464-000, Tucuru\'{\i}, Par\'{a}, Brazil}

\affiliation{$^{(d)}$ Universidade Federal Rural da Amaz\^{o}nia ICIBE - 
LASIC\\
Av. Presidente Tancredo Neves 2501 CEP 66077-901 - Bel\'{e}m/PA, Brazil}

\affiliation{$^{(e)}$ Universidade Federal do Esp\'{\i}rito Santo, CEP 29075-910, Vit\'{o}ria, ES, Brazil}

\affiliation{$^{(f)}$ National Research Nuclear University “MEPhI”, Kashirskoe sh. 31, Moscow 115409, Russia}

\begin{abstract}

In this paper, we determine regular black hole solutions using a very general $f(R)$ theory, coupled to
a non-linear electromagnetic field given by a Lagrangian $\mathcal{L}_{NED}$. The functions $f(R)$ and $\mathcal{L}_{NED}$ 
are left in principle unspecified. Instead, the model is constructed through a choice of the mass function $M(r)$ presented in the metric coefficients. Solutions which have a regular behaviour of the geometric invariants are found. These solutions have two horizons, the event horizon and the Cauchy horizon. All energy conditions are satisfied in the whole space-time, except the strong energy condition (SEC) which is violated near the Cauchy horizon.

\end{abstract}
\pacs{ 04.50.Kd}

\maketitle


\section{Introduction}
\label{sec1}

The present stage of accelerate expansion of the universe seems to be well established from the analysis of observational data. Besides the supernova Ia data \cite{SNIa}, the data from the observation of the anisotropy of the cosmic microwave background radiation (CMB) \cite{CMB}, the baryonic  acoustic oscillations (BAO) \cite{BAO}, large scale structures \cite{LSS},weak lensing \cite{WL}, the differential age of old galaxies ($H(z0$) \cite{h}, give strong evidences for the present accelerated expansion phase. Since gravity is attractive, the cosmic accelerate expansion requires some new form of exotic matter that leads to violation of the Strong Energy Condition (SEC) \cite{sing-theo,h-e}, as far as the General Relativity (GR) theory is considered. This exotic component is dubbed dark energy. 

The most popular, and most simple, candidate for dark energy is the cosmological constant.
Interpreted as a manifestation of the quantum vacuum energy, the cosmological constant faces however a huge discrepancy between the observed value and the predicted one. The exact value of this discrepancy depends on many details, but in general it mounts to many dozen orders of magnitude \cite{weinberg}.

The incertitude about the dynamical origin of the observed accelerated expansion led to many speculations about
possible extensions of the General Relativity (GR) theory in such way that the accelerated expansion could be obtained without the introduction of dark energy. In this sense, one of these possible extensions is to generalise the Einstein-Hilbert action including non-linear geometric terms. One of this proposal is the $f(R)$ theories \cite{fR,deFelice}, where the non-linear terms
are combinations of the Ricci scalar $R$. Such theories may give very good results at cosmological scales but must be complemented with a screening mechanism in order not to spoil the achievements of the GR theory at scales of the solar system \cite{koyama}. There are a long list of other possible, and generally more complex, modifications of the GR theory \cite{fRT,fG,fRG,TT,fT,fTTc,fTTG}

Another problem concerning the applications of the GR theory to concrete problems is the presence of singularities, as that predicted in the primordial universe and in the end of the life of some massive stars. The presence of such singularities seems to point to the limit of application of the GR theory, requiring perhaps to consider quantum effects in the strong gravitational regime. Some other possibility to cure this singularity problem, yet in the context of a classical theory, is to consider as source of the gravitational equations matter fields that may lead to violation of at least some of the energy conditions. Examples are given by non-linear gauge fields, like the electromagnetic field. Non linear electromagnetism \cite{BI} has been conceived originally to cure singularity problems in the Maxwell theory. In gravity theories, the electromagnetic field appears as one of the sources of of the structure of the space-time. In such a context, some success to avoid singularities has been obtained in implementing such extension of the classical Maxwell field \cite{peres,NED,Bardeen,regulardeSitter,ayon-BG_Br}.

In the Ref. \cite{zanchin} both proposals of extension of the usual gravitational and gauge field theories were considered.
In that paper, the emphasis was in the study of black hole configurations. A very general form of the this theory mixing 
the $f(R)$ theory and the non-linear electromagnetic Lagrangian $\mathcal{L}_{NED}$ has been considered. A static, spherically symmetric space-time has been used. Particular, solutions with horizon (thus, candidates to represent a black hole) were found, but
without the singularity existing in the usual black hole solutions of the GR theory. This singularity-free black hole solutions imply the violation of the SEC only in certain regions of the space-time. However, the other energy conditions 
are generally satisfied. We remember that the energy conditions are directly connected with the existence of singularities in GR theory \cite{sing-theo,h-e}.

In the present paper, we revisit the problem treated in the Ref. \cite{zanchin}, and we show that new non-singular solutions are possible.  This new solutions emerge from a specific, but very appealing, choice of the mass function $M(r)$, which will be properly defined later. The mass function we will use was constructed in Ref. \cite{balart1}, with GR theory coupled to non-linear electromagnetic field, in other to satisfy some requirements, like to avoid violation of the Weak Energy Condition (WEC) and to have the Reissner-Nordstr\"om asymptotic limit:  the mass function $M(r)$ given in Ref. \cite{balart1} is the most general functional form satisfying the WEC in GR.  As it was found for other mass functions in Ref. \cite{zanchin}, the employment of
the mass function of Ref. \cite{balart1} in our general context implies that the violation of SEC occurs only in a limited
region of the space-time, the other energy conditions being satisfied in the entire space-time.

This paper is organised as follows. In the next section, the equations of motion are written down. In section III, we determine the new non-singular solutions and analyse the fate of the energy conditions for these solutions. 
The final conclusions are presented in section IV. In the appendix, it is shown explicitly that the solutions found here
are asymptotically regular.


\section{The equations of motion in $f(R)$ Gravity}
\label{sec2}

The $f(R)$ gravity is defined by the action,
\begin{eqnarray}
S_{f(R)}=\int d^4x\sqrt{-g}\left[f(R)+2\kappa^2\mathcal{L}_m\right]
\label{action}\;,
\end{eqnarray}
where $g$ stands for the determinant of the metric $g_{\mu\nu}$, $f(R)$ is a 
given function of the Ricci scalar $R$, $\mathcal{L}_m$ represents the 
Lagrangian density of the matter and other fields, and 
$\kappa^2=8\pi G/c^4$, with $G$ and $c$ being the Newton's 
gravitational constant and the speed of light, respectively. 

There are two main approaches for this theory, the first one supposing the the dynamical fields are the metric and the matter field, known as metric formalism,
and the second one, called Palatini formalism, for which the dynamical fields are the metric, the matter field, but with the Levi-Civita connection independent of the metric. In what follows we will use the first approach. 

Applying the variational principle in terms of the metric to the action (\ref{action}), we find the following field equations: 
\begin{eqnarray}
f_RR^{\mu}_{\;\;\nu}-\frac{1}{2}\delta^{\mu}_{\nu}f+\left(\delta^{\mu}_{\nu} 
\square 
-g^{\mu\beta}\nabla_{\beta}\nabla_{\nu}\right)f_R=\kappa^2\Theta^{\mu}_{
\;\;\nu}\label{eqfR}\;,
\end{eqnarray}  
where $f_R\equiv df(R)/dR$, $R^{\mu}_{\;\;\nu}$  is the Ricci 
tensor, $\nabla_{\nu}$ stands for the covariant 
derivative, $\square\equiv 
g^{\alpha\beta}\nabla_{\alpha}\nabla_{\beta}$ is the d'Alembertian, and 
$\Theta_{\mu\nu}$ is the matter energy-momentum tensor. 
\par 
In the present work, we will analyse the coupling of the $f(R)$ gravity with a Non-linear electrodynamic theory (NED), given by, $\mathcal{L}_m\equiv 
\mathcal{L}_{NED}(F)$, where $F=(1/4)F^{\mu\nu}F_{\mu\nu}$, and with  
$F_{\mu\nu}$ being the Maxwell tensor, and $
\mathcal{L}_{NED}(F)$ is an arbitrary function of $F$.   A similar structure was exploited in Ref. \cite{zanchin}. We will first review the methodology employed in that reference, which will be applied in the present paper in order to find new regular black hole solutions.
\par
Considering the NED coupling, the energy-momentum tensor for matter in (\ref{eqfR}) is given by,
\begin{eqnarray}
&&\Theta^{\mu}_{\;\;\nu}=\delta^{\mu}_{\nu}\mathcal{L}_{NED}-\frac{
\partial\mathcal{L}_{NED}(F)}
{\partial F}F^{\mu\alpha}F_{\nu\alpha}\label{energy}\;.
\end{eqnarray}
In the particular case of the Maxwell Lagrangian $\mathcal{L}_{NED}\equiv F$, the energy-momentum tensor of the linear Maxwell electrodynamics is
reobtained.
\par 
Defining the Maxwell tensor in terms of the four-potential $F_{\mu\nu}=\partial_{\mu}A_{\nu}-\partial_{\nu}A_{\mu}$, the variation of the functional (\ref{action}) with respect to the potential can be performed, leading to the generalised Maxwell equation,
\begin{eqnarray}
\nabla_{\mu}\left[F^{\mu\nu}\mathcal{L}_F\right]\equiv  
\partial_{\mu}\left[\sqrt{-g}F^{\mu\nu}\mathcal{L}_F\right]=0
\label{Maxwell}\;,
\end{eqnarray}
where $\mathcal{L}_F=\partial\mathcal{L}_{NED}/\partial F$.
\par 
We then consider a spherically symmetric and static space-time, whose element 
line, in Schwarzschild coordinates, reads
\begin{eqnarray}
ds^2=e^{a(r)}dt^2-e^{b(r)}dr^2-r^2\left[d\theta^2+\sin^2\theta 
d\phi^2\right] 
\label{ele}\;,
\end{eqnarray} 
where $a(r)$ and $b(r)$ are arbitrary functions of the radial coordinate 
$r$. 
We will consider the particular case where there is only electric field, the components connected with the magnetic field of the Maxwell tensor $F_{\mu\nu}$ being zero. Imposing spherical symmetry, through the Killing vectors and the equation $\mathcal{L}_{\zeta^{\mu}}F^{\alpha\beta}(t,r,\theta,\phi)\equiv 0$, we can show that only non-null component of the Maxwell tensor is $F^{10}(r)$ \cite{wainwright}.  The generalised Maxwell equation  \eqref{Maxwell} for $\nu =0$ is 
\begin{eqnarray}
F^{10}(r)=\frac{q}{r^2}e^{-[a(r)+b(r)]/2}\mathcal{L}_F^{-1}
(r) \label {F10-1} \; ,
\end{eqnarray} 
where $q\in\Re$ is an integration constant representing the electric 
charge of the source.
\par 
The equations of motion for the $f(R)$ Gravity coupled to a NED are then 
found by using the line element \eqref{ele}, the energy-momentum \eqref{energy}, with the only non-null component given by
\eqref{F10-1}, and the field equations
\eqref{eqfR}:
\begin{eqnarray}
&&\frac{e^{-b}}{4r}\Big\{4r\frac{d^2f_R}{dr^2}+ 
2\left[4-rb'\right]\frac{df_R}{dr}+\big[ra'b'-2ra''-r(a')^2-4a'\big]f_R+2re^bf\Big\}=-\kappa^2\left[\mathcal{L}_{NED}+ \frac{q^2}{r^4}\mathcal{L}_F^{-1}\right] ,\label{eq1}\\
&& \frac{e^{-b}}{4r}\Big\{2\left[4+ra'\right]\frac{df_R}{dr}+ 
\left[(4+ra')b'-2ra''-r(a')^2\right]f_R+2re^bf\Big\}=-\kappa^2\left[\mathcal{L}_{NED}+
\frac{q^2}{r^4}\mathcal{L}_F^{-1}\right],\label{eq2}\\
&&\frac{e^{-b}}{2r^2}\Big\{2r^2\frac{d^2f_R}{dr^2}+[r^2(a'-b')+2r]
\frac{df_R}{dr}+[r(b'-a')+2(e^b-1)]f_R+r^2e^bf\Big\} =-\kappa^2\mathcal{L}_{NED} ,\label{eq3}
\end{eqnarray}
where the prime ($'$) stands for the total derivative with respect to the 
radial coordinate $r$. 
\par
In the next section, we will use an algebraic methodology to solve these equations and to obtain new regular solutions.

\section{New generalizations for regular black holes on General Relativity to $f(R)$ Gravity}\label{sec3}
Fixed the spherical coordinates, it is possible to impose the quasi-global condition by choosing a radial coordinate: 
\begin{eqnarray}
b(r)=-a(r)\label{b}\;.
\end{eqnarray}
This is an additional requirement to the metric functions since the coordinate system has already been fixed. However, the fact that the functions $f(R)$ and $\mathcal{L}_{NED}$ are, for the moment, arbitrary assures the possibility to impose such new condition, as it will be verified later.

Imposing the quasi-global coordinate condition, and combining equations \eqref{eq2} and \eqref{eq1}, we obtain, 
\begin{eqnarray}
e^{-b}\frac{d^2f_R}{dr^2}=0\label{eq4}\;.
\end{eqnarray}
Integrating this expression, it results,
\begin{eqnarray}
f_R(r)=c_1r+c_0\label{fR1}\;,
\end{eqnarray}
where it appears the integration constants $c_0,c_1\in\Re$. Here, we must make the following observations. First, for the particular case $c_1\equiv 0,c_0=1$, GR is recovered, since the integration of \eqref{fR1} with respect to $R$ leads to $f(R)=R$. Second, if the line element \eqref{ele} is considered, the Ricci scalar becomes, 
\begin{eqnarray}
R &=& e^{-b}\left[a''+\left(a'-b'\right)\left( \frac{a'}{2} +\frac{2}{r}\right)
+\frac{ 2}{r^2}  \right]-\frac{2}{r^2}\;\nonumber\\
&=& e^{a}\left[a''+ 2a'\left( \frac{a'}{2} +\frac{2}{r}\right)
+\frac{ 2}{r^2}  \right]-\frac{2}{r^2}\label{R}\;.
\end{eqnarray}
The regularity of the solution (\ref{fR1}) at the spatial infinity is discussed in the appendix. 

In order to have a better description of the new regular black hole solutions, it is useful to define,
\begin{eqnarray}
e^{a(r)}=1-\frac{2M(r)}{r}\label{a}\;,
\end{eqnarray}
where $M(r)$ is the mass function, which for the regular solutions satisfies the condition $\lim_{r\rightarrow 0}[M(r)/r]\equiv 0$. The mass function $M(r)$ must coincide with the ADM mass $m$ in the spatial infinity limit, where the radial coordinate $r$ goes to infinity. 
\par 
Inserting \eqref{a} and \eqref{b} in \eqref{R}, the curvature scalar can be rewritten as,
\begin{eqnarray}
R(r)=-\frac{2}{r^2}[rM''(r)+2M'(r)]\label{R1}\;,
\end{eqnarray}
which, for a given $M(r)$ model, may allow to invert equation \eqref{R1}, to obtain $r(R)$. After integration of \eqref{fR1}
it leads to,
\begin{eqnarray}
f(R)=c_0R+c_1\int r(R)dR\;.\label{f1}
\end{eqnarray}
It is also possible to obtain $f(R)$ from the expression $f_R=(df/dr)(dR/dr)^{-1}$. After integration and using \eqref{R1} we obtain,
\begin{eqnarray}
f(r)=\int f_R(r)\frac{dR(r)}{dr}dr\label{f2}\;.
\end{eqnarray} 
Now, we use the methodology presented in \cite{zanchin} to solve the equations of motion.
Taking the relations \eqref{b}, \eqref{a}, \eqref{fR1} and \eqref{f2}, we can solve the equations \eqref{eq1}-\eqref{eq3} to obtain $\mathcal{L}_{NED}$ and $\mathcal{L}_F$ as,
\begin{eqnarray}
&&\mathcal{L}_{NED}=-\frac{1}{2\kappa^2 r^2}\left[r^2f(r) 
+4c_0M'(r)+2c_1r\right]\label{L}\;,\\
&&\mathcal{L}_F=-\kappa^2 \frac{q^2}{r^2}\big[(c_1r+c_0)rM''(r) 
-(2c_0+c_1r)M'(r)-3c_1M(r)+c_1r\big]^{-1}\label{LF}\;.
\end{eqnarray}

The solution above satisfies the equations of motion. However, its consistency can be verified through the Lagrangian density $\mathcal{L}_{NED}$ and its derivative with respect to $F$, $\mathcal{L}_F$. By definition, we have
\begin{eqnarray}
\mathcal{L}_F=\frac{\partial \mathcal{L}_{NED}}{\partial F}=\frac{\partial 
\mathcal{L}_{NED}}{\partial r}\frac{\partial r}{\partial F}=\frac{\partial 
\mathcal{L}_{NED}}{\partial r}\left(\frac{\partial F}{\partial 
r}\right)^{-1}\;.\label{Lconstraint}
\end{eqnarray}
To perform such verification we must remember that $F=(1/4)F^{\mu\nu}F_{\mu\nu}=-(1/2)e^{a+b}[F^{10}(r)]^2$, and that the only non-null component of the Maxwell's tensor, for this symmetry, is obtained from equations  \eqref{b}, \eqref{a} and \eqref{LF}, which considering \eqref{F10-1}, leads to
\begin{eqnarray}
&&F^{10}(r)=\frac{1}{q\kappa^2}\Big\{3c_1M(r)+(2c_0+c_1r)M'(r)-r[c_1+(c_0+c_1r)M''(r)]\Big\}\label{F10-2}\;.
\end{eqnarray}
Now, using equations \eqref{b}, \eqref{fR1}, \eqref{a}, \eqref{R1}, \eqref{f2}-\eqref{LF} and \eqref{F10-2}, it is possible to verify that the constraint \eqref{Lconstraint} is satisfied. Hence, the solution is consistent. 
\par 
In order to perform an analysis of the physical properties of this class of solution, we must take into account the energy condition relations for the $f(R)$ theory. Following the results of Refs. \cite{santos,visser}, equation \eqref{eqfR} is rewritten as,
\begin{eqnarray}
&&R_{\mu\nu}-\frac{1}{2}g_{\mu\nu}R=f_R^{-1}\big[\kappa^2\Theta_{\mu\nu}
+\frac{1}{2}g_{\mu\nu}\left(f-Rf_R\right)-\left(g_{\mu\nu}\square-\nabla_{\mu}\nabla_{\nu}\right)f_R\big]=\kappa^{2
}  \mathcal{T}_{\mu\nu}^{(eff)}\label{energyeff}\;,
\end{eqnarray}
where $\mathcal{T}^{(eff)}_{\mu\nu}$ is the effect if energy-momentum tensor, and the perfect fluid content is identified by the relations  $\mathcal{T}^{0(eff)}_{0}=\rho^{(eff)},\mathcal{T}^{1(eff)}_{1}=-p^{(eff)}_r,\mathcal{T}^{2(eff)}_{2}=\mathcal{T}^{3(eff)}_{3}=-p^{(eff)}_t$, where $\rho^{(eff)}$, $p^{(eff)}_r$ e $p^{(eff)}_t$ are the energy density, radial and tangential pressures, respectively. The explicit expressions for the energy-momentum tensor can be found in Ref.\cite{zanchin}. With these expressions, the energy conditions for the $f(R)$ theory can be written as, 
\begin{eqnarray}
&&NEC_{1,2}(r)=\rho^{(eff)}+p_{r,t}^{(eff)}\geq 0\;,\label{cond1}\\
&&SEC(r)=\rho_{(eff)}+p_{r}^{(eff)}+2p_{t}^{(eff)}\geq 0\,,\label{sec}\\
&&WEC_{1,2}(r)=\rho^{(eff)}+p_{r,t}^{(eff)}\geq 0\;,\label{cond2}\\
&& DEC_{1}(r)=\rho^{(eff)}\geq 0, \label{cond3a}\\ &&
DEC_{2,3}(r)=\rho^{(eff)}-p_{r,t}^{(eff)}\geq 
0\;,\label{cond3}
\end{eqnarray}
where, in view of the identity $WEC_3(r)\equiv DEC_1(r)$, one of the conditions was not written.
\par 
In the next sub-section, we will use a specific model for the general mass function
$M(r)$, coming from GR, in order to obtain a generalisation of this class of solutions. 

\subsection{New regular black hole solutions}
As it has been shown in reference \cite{zanchin}, it is not any model for the mass function $M(r)$ that leads to a generalisation of a solution of GR to the $f(R)$ gravity theory, with a function $f(R)$ containing non-linear terms. Here, we will use a model obtained by integration of the general mass function satisfying the WEC given by \cite{balart1}, which reads,
\begin{eqnarray}
&&M(r)=\frac{6^3m^4r^3}{q^6}\frac{\Gamma^3(4/a_1)\Gamma(b_1/a_1)}{\Gamma^3(a_1^{-1})\Gamma[(b_1-3)/a_1]}\,_{2}F_{1}\Big[\frac{3}{a_1};\frac{b_1}{a_1};\frac{a_1+3}{a_1};-\left(\frac{6\Gamma(4/a_1)}{\Gamma(a_1^{-1})\Gamma[(a_1+3)/a_1]}\frac{m}{q^2}r\right)^{a_1}\Big]\label{M1}\,,
\end{eqnarray} 
where $_2F_1[k_1;k_2;k_3;z]$ is the Gauss hypergeometric function. 
\par 
This model is, in general, very complicated. It is possible to express the solutions in terms of integrals, which can be or not analytical. It is more instructive to work with some particular cases. Let us first take the case where $a_1=2$ and $b_1=4$,  for which the metric functions \eqref{b} and \eqref{a} are given by,
\begin{eqnarray}
e^{a}=e^{-b}=1+\frac{32m^2q^2}{\pi^2q^4+64m^2r^2}-\frac{4m}{\pi r}\arctan\left[\frac{8mr}{\pi q^2}\right]\label{a2}\;.
\end{eqnarray}  
This solution represents a charged, regular black hole, asymptotically flat, with two horizons: $r_H$ (an event horizon) and $r_-$ (inner or Cauchy horizon). It is possible to verify that this solution is regular in all space-time by inspecting the Ricci and Kretschmann scalar, which read,
\begin{eqnarray}
&&R=-\frac{16384m^4\pi^2q^6}{(\pi^2q^4+64m^2r^2)^3}\label{R2}\;,
\end{eqnarray}
\begin{eqnarray}
&&\mathcal{K}=R^{\alpha\beta\mu\nu}R_{\alpha\beta\mu\nu}=\frac{64m^2}{r^6}\Big\{\frac{64m^2q^4r^2}{(\pi^2q^4+64m^2r^2)^6}(3\pi^4q^8+256\pi^2m^2q^4r^2+20480m^4r^4)(\pi^4q^8+256\pi^2m^2q^4r^2+28672m^4r^4)\nonumber\\
&&+\frac{1}{\pi^2}\arctan\left(\frac{8mr}{\pi q^2}\right)\Big[3\arctan\left(\frac{8mr}{\pi q^2}\right)-\frac{16m\pi q^2r}{(\pi^2q^4+64m^2r^2)^3}(3\pi^4q^8+512m^2\pi^2q^4r^2+36864m^4r^4)\Big]\Big\}\label{kre1}.
\end{eqnarray}
These scalars are finite in all space-time. The limits in the origin of the radial coordinate and in the spatial infinity are given by $\lim_{r\rightarrow 0}\{R,\mathcal{K}\}=\{-(16384m^4)/(\pi^4q^6),(134217728m^8)/(3\pi^8q^{12})\}$ and $\lim_{r\rightarrow \infty}\{R,\mathcal{K}\}=\{0,0\}$.
\par 
The expression \eqref{R2} implies,
\begin{eqnarray}
r(R)=\frac{q}{8m}\sqrt{-\pi^2q^2+\frac{16(2\pi m^2)^{2/3}}{(-R)^{1/3}}}\label{rR1}\;.
\end{eqnarray}
From \eqref{R2} it can be verified that $R\leq 0$, what must be taken into account in order to define correctly \eqref{rR1} in the usual limit $0\leq r\leq +\infty$.
\par
Now, the function $f(R)$ can be obtained inserting \eqref{rR1} in \eqref{f1}, leading to,
\begin{eqnarray}
&&f(R)=c_0R+\frac{c_1}{8mq^3\pi^{8/3}}\sqrt{-\pi ^2 q^2+\frac{16(2\pi m^2)^{2/3}}{(-R)^{1/3}}}\big[192(2m^8)^{1/3}(-R)^{1/3}+4(4\pi^{4}m^4)^{1/3}(-R)^{2/3}-\pi^{8/3}q^4 (-R)\big]\nonumber\\
&&+\frac{768c_1m^3}{\pi ^3q^4}\arctan^{-1}\Big\{\pi q\Big[-\pi ^2q^2+\frac{16(2\pi m^2)^{2/3}}{(-R)^{1/3}}\Big]^{-1/2}\Big\}
\end{eqnarray}
We can now see clearly that the last two terms that multiply the constant $c_1$ generalize the GR solution to  the $f(R)$ gravity, including non-linear terms in the Ricci scalar $R$. In the particular case where $c_1=0$,  GR is recovered. 
\par
\begin{figure}[h]
\centering
\begin{tabular}{rl}
\includegraphics[height=5cm,width=8cm]{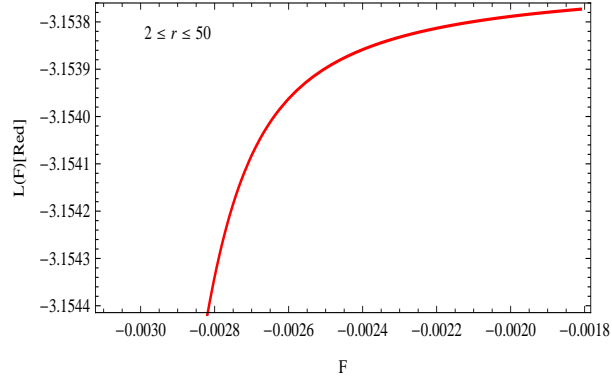}
\end{tabular}
\caption{\scriptsize{Parametric representation $\{\mathcal{L}_{NED},F\}$ of the solution \eqref{a2}, with $q=10, m=80,c_0=1, c_1=2, \kappa^2=8\pi$.} }
\label{fig1}
\end{figure}

 
It is not possible to obtain analytically a functional relation between the scalar $F$ and the Lagrangian density $\mathcal{L}_{NED}$. Hence, using \eqref{F10-2}, with \eqref{M1}, we obtain
\begin{eqnarray}
&&F^{10}(r)=\frac{1}{\pi q\kappa^2(\pi^2q^4+64m^2r^2)^3}\Bigg\{-\pi r\Big[-524288c_0m^6q^2r^3+c_1\Big(\pi^6q^{12}+4096m^4\pi^2q^4r^2
(2q^2+3r^2)+65536m^6r^4\times\nonumber\\
&&\times(-3q^2+4r^2)+48m^2\pi^4q^8(q^2+4r^2)\Big)\Big]+6c_1m(\pi^2q^4+64m^2r^2)^3\arctan\left(\frac{8mr}{\pi q^2}\right)\Bigg\}
\end{eqnarray}
Now, $\mathcal{L}_{NED}$ can be obtained through \eqref{L}:
\begin{eqnarray}
&&\mathcal{L}_{NED}(r)=-\frac{1}{\kappa^2}\Bigg\{\frac{c_1}{r}-\frac{8192m^4\pi^2q^6}{(\pi^2q^4+64m^2r^2)^3}(c_0+c_1r)+\frac{2048m^4q^2}{(\pi^2q^4+64m^2r^2)^2}(2c_0+c_1r)+\frac{3072c_1m^4r}{\pi^2q^2(\pi^2q^4+64m^2r^2)}\nonumber\\
&&+\frac{384c_1m^3}{\pi^3q^4}\arctan\left(\frac{8mr}{\pi q^2}\right)\Bigg\}
\end{eqnarray}
It is possible to represent parametrically a graphic $\mathcal{L}_{NED}(F)\times F$, where $F=(1/4)F^{\mu\nu}F_{\mu\nu}$. This behaviour is displayed in figure \ref{fig1}.

The energy conditions can now be verified. Taking explicitly the effective density and pressure \cite{zanchin} for our particular case \eqref{a2}, we find the following expressions for the energy conditions:
\begin{eqnarray}
&&NEC_1(r)=WEC_1(r)=0, NEC_2(r)=WEC_2(r)=\frac{524288m^6q^2r^2}{\kappa^2(\pi ^2q^4+64m^2r^2)^3},\\
&&DEC_1(r)=\frac{1}{2}DEC_2(r)=\frac{4096m^4q^2}{\kappa^2(\pi ^2q^4+64m^2r^2)^2},DEC_3(r)=\frac{8192m^4\pi^2q^6}{\kappa^2(\pi ^2q^4+64m^2r^2)^3}\\
&&SEC(r)=\frac{8192m^4q^2[(8mr)^2-\pi ^2q^4]}{\kappa^2(\pi ^2q^4+64m^2r^2)^3}\label{sec1}
\end{eqnarray} 
The energy conditions NEC, WEC and DEC are satisfied in all space-time. However, the SEC energy condition \eqref{sec1}, is violated for $r<[\pi q^2/(8m)]$ which, for the values $m=8q,q=10$, represents a region very near the Cauchy horizon ($\pi q^2/(8m)=0.490874,r_{Cauchy}=0.0419174,r_{H}=159.373$). 
\par 
Let us show a second analytical example for the general mass \eqref{M1}. For the values $a_1=3$ and $b_1=4$, using the same procedure as before, the following functions characterise the solutions:
\begin{eqnarray}
&&e^{a(r)}=e^{-b(r)}=1-\frac{2m}{r}\left[1-\frac{q^2}{(q^6+8m^3r^3)^{1/3}}\right],R(r)=-\frac{64m^4q^8}{(q^6+8m^3r^3)^{7/3}}\label{a3}\\
&&F^{10}(r)=\frac{256c_0m^7q^2r^5}{\kappa^2 q(q^6+8m^3r^3)^{7/3}}+\frac{c_1}{\kappa^2 q}\left[3m-r+\frac{256m^7q^2r^6}{(q^6+8m^3r^3)^{7/3}}-\frac{8m^4q^2r^3}{(q^6+8m^3r^3)^{4/3}}-\frac{3mq^2}{(q^6+8m^3r^3)^{1/3}}\right]\label{F10-3}\\
&&
\mathcal{K}=\frac{16m^2}{r^6(q^6+8m^3r^3)^{14/3}}\Big[3q^{28}+112m^3q^{22}r^3+1856m^6q^{16}r^6+14336m^9q^{10}r^9+57344m^{12}q^4r^{12}-(q^6+8m^3r^3)^{1/3}\times \nonumber\\
&&\times\left(6q^{26}+208m^3q^{20}r^3+2944m^6q^{14}r^6+19456m^9q^8r^9+49152m^{12}q^2r^{12}\right)+(q^6+8m^3r^3)^{2/3}\Big(3q^{24}+96m^3q^{18}r^3\nonumber\\
&&+1152m^6q^{12}r^6+6144m^9q^6r^9+12288m^{12}r^{12}\Big)\Big]\label{kre3}\\
&&f(R)=c_0R+\frac{c_1}{2mq^{12}(mq^2)^{16/3}}(-R)^{16/21}[4\times 2^{4/7}(mq^2)^{12/7}-q^6(-R)^{3/7}]^{4/3}[3\times 2^{4/7}(mq^2)^{12/7}(-R)^{4/7}-q^6R]\label{f3}\,.
\end{eqnarray}

In the the spatial infinity, we have $\lim_{r\rightarrow +\infty}\{e^{a(r)},e^{b(r)}\}=\{1,1\}$ and $\lim_{r\rightarrow +\infty}\{R(r),\mathcal{K}\}=\{0,0\}$, showing the regularity in the asymptotical region. When the radial coordinate goes to zero it is necessary to perform an expansion around $r = 0$ to analyse the behaviour of the solution in this limit. Taking $e^{a(r)}$ and $R(r)$ , \eqref{a3}, $\mathcal{K}$ inserting in\eqref{kre3}, we find the following expansions around $r=0$:
\begin{eqnarray}
&&e^{a(r)}\sim 1-\frac{16m^4}{3q^{6}}r^2+O\left(r^3\right),R(r)\sim -\frac{64m^4}{q^{6}}+O\left(r^3\right),\mathcal{K}\sim \frac{2048m^8}{3q^{12}}+O\left(r^3\right)\label{a4}
\end{eqnarray}
When $r$ goes to zero, we have $\lim_{r\rightarrow 0}\{e^{a(r)},e^{b(r)}\}=\{1,1\}$ e $\lim_{r\rightarrow 0}\{R(r),\mathcal{K}\}=\{-64m^4/q^{6},2048m^8/q^{12}\}$, showing the regularity in the origin of the radial coordinate. Hence, we have shown that the solution \eqref{a3} corresponds to a regular black hole in all space-time. This black hole has spherical symmetry, and it is charged and asymptotically flat.
\par 
Let us verify now the energy conditions for this case. They read,
\begin{eqnarray}
&&NEC_1(r)=WEC_1(r)=0, NEC_2(r)=WEC_2(r)=\frac{256m^7q^2r^3}{\kappa^2(q^6+8m^3r^3)^{7/3}},\\
&&DEC_1(r)=\frac{1}{2}DEC_2(r)=\frac{16m^4q^2}{\kappa^2(q^6+8m^3r^3)^{4/3}},DEC_3(r)=\frac{32m^4q^8}{\kappa^2(q^6+8m^3r^3)^{7/3}}\\
&&SEC(r)=\frac{32m^4q^2(8m^3r^3-q^6)}{\kappa^2(q^6+8m^3r^3)^{7/3}}\label{sec2}
\end{eqnarray} 
The conditions NEC, WEC and DEC are satisfied in all space-time. However, the SEC condition, given by \eqref{sec2}, is violated for $r<q^2/(2m)$. For the values $m=8q,q=10$, this violation occurs in region very near the Cauchy horizon ($q^2/(2m)=0.625,r_{Cauchy}=0.0676869,r_{H}=159.373$). 
\par
\begin{figure}[h]
\centering
\begin{tabular}{rl}
\includegraphics[height=5cm,width=8cm]{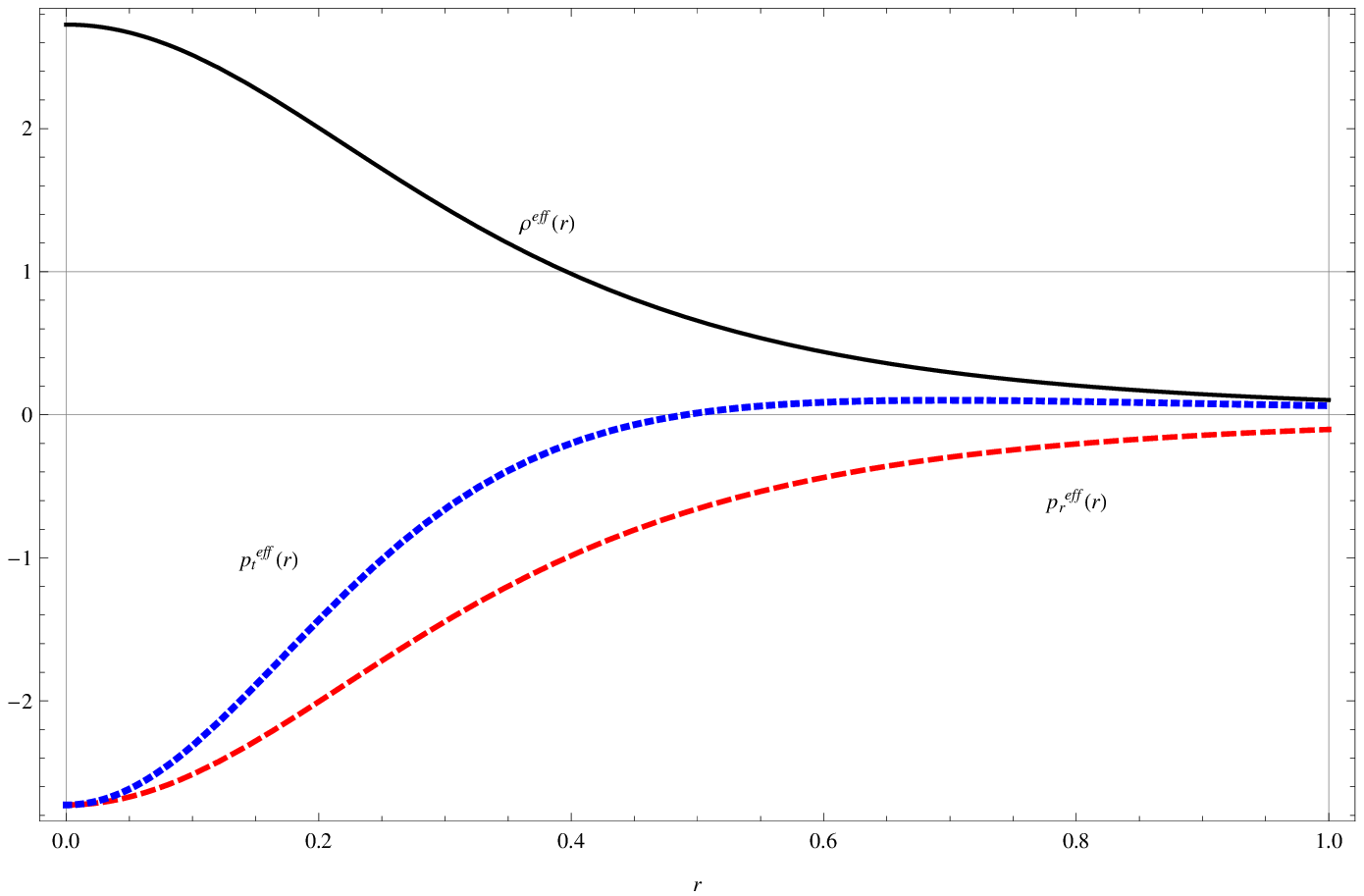}&\includegraphics[height=5cm,width=8cm]{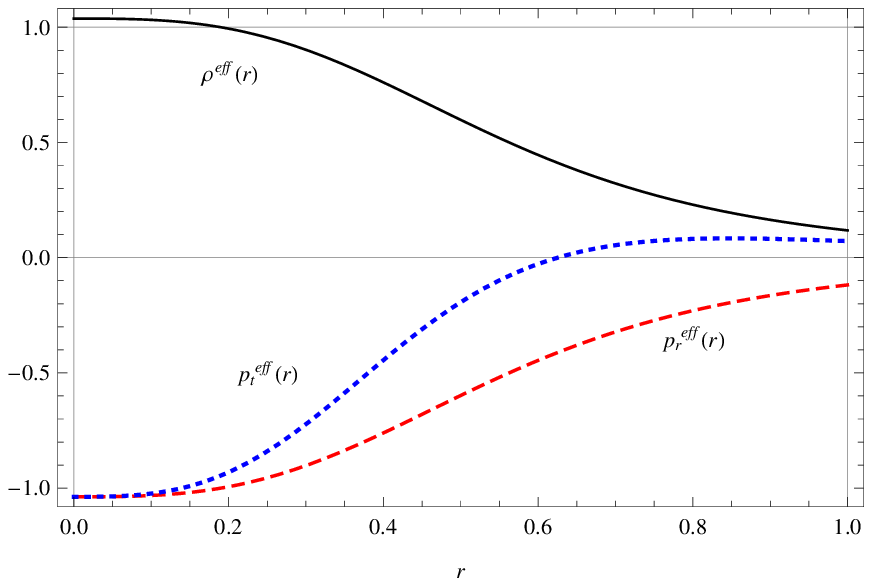}\\
\includegraphics[height=5cm,width=8cm]{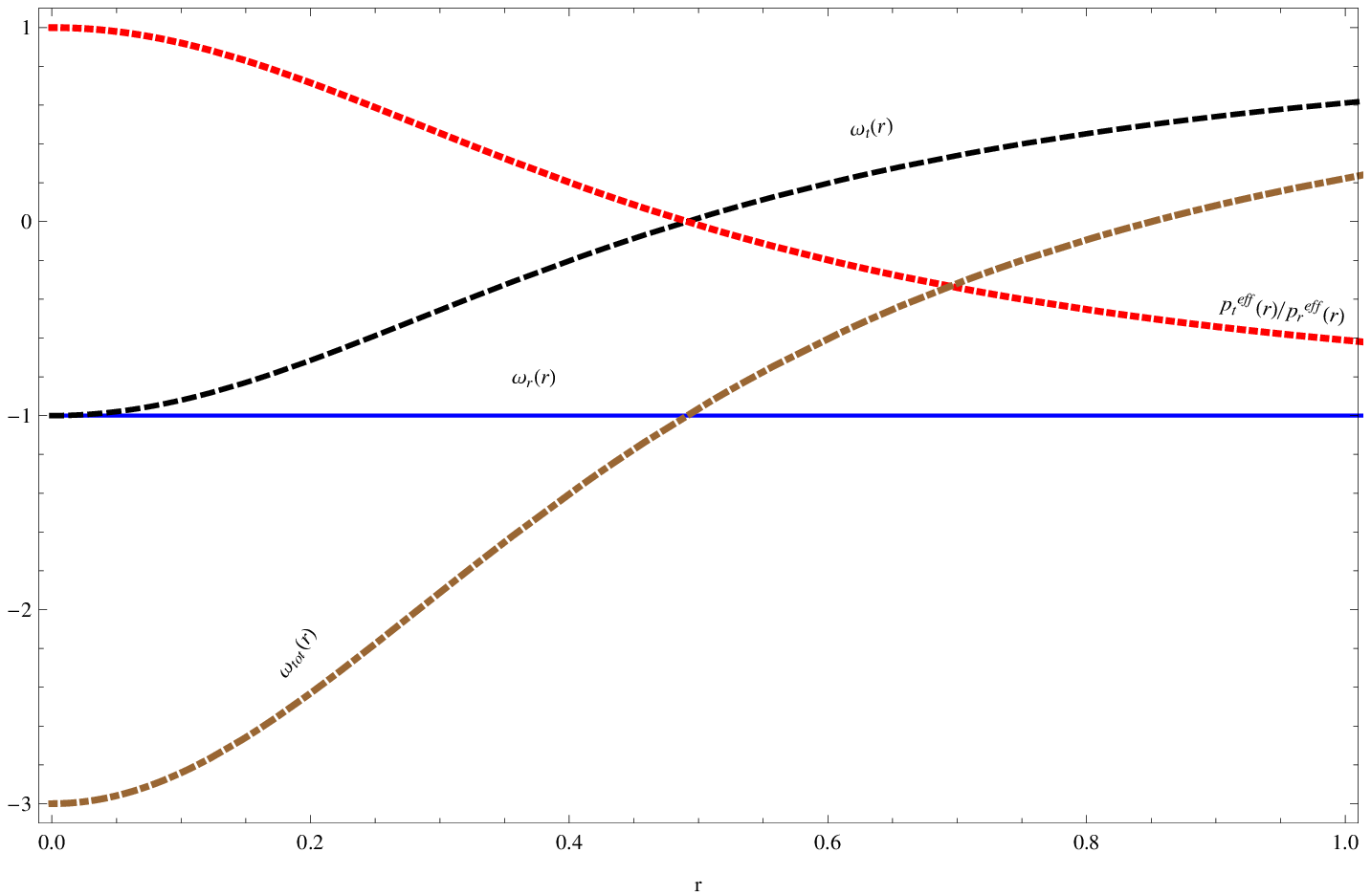}&\includegraphics[height=5cm,width=8cm]{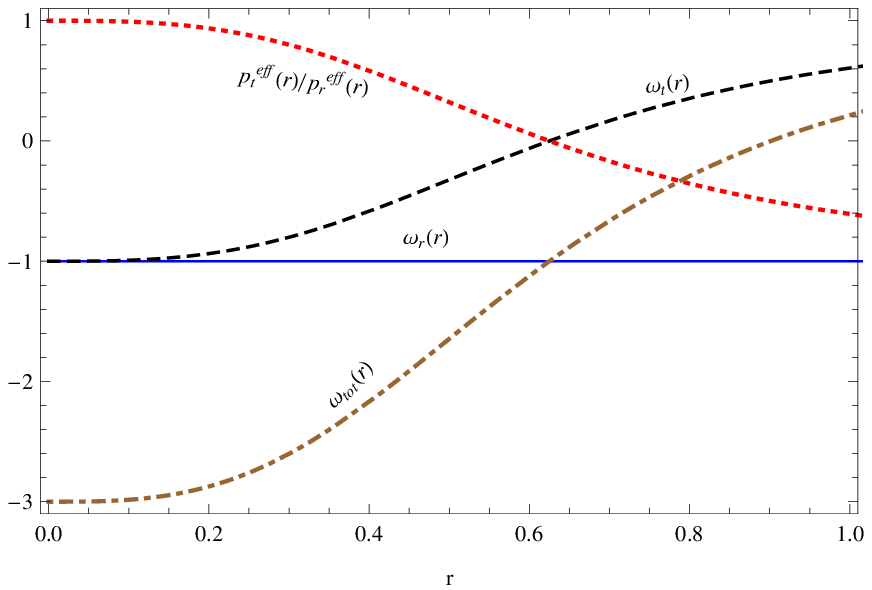}
\end{tabular}
\caption{\scriptsize{Parametric representation of the density and effective pressures of the solution \eqref{a2} (left up panel) and of the solution \eqref{a3} (right up panel). It is also displayed the factions $\omega_r=p_{r}^{(eff)}/\rho^{(eff)},\omega_t=p_{t}^{(eff)}/\rho^{(eff)},\omega_{eff}=(p_{r}^{(eff)}+2p_{r}^{(eff)})/\rho^{(eff)}$ and $p_{r}^{(eff)}/p_{t}^{(eff)}$. We used $q=10, m=80, c_0=1, c_1=2, \kappa^2=8\pi$.}} 
\label{fig2}
\end{figure}

 
To complete our analysis, we display in figure \ref{fig2} the graphics for the density, the radial and tangential effective pressures for the solutions \eqref{a2} and \eqref{a3}. We show also the graphics for the relative fractions of these quantities, $\omega_r=p_{r}^{(eff)}/\rho^{(eff)},\omega_t=p_{t}^{(eff)}/\rho^{(eff)},\omega_{eff}=(p_{r}^{(eff)}+2p_{r}^{(eff)})/\rho^{(eff)}$ and $p_{r}^{(eff)}/p_{t}^{(eff)}$.   
It can be seen from figure \ref{fig2} that the radial pressure reveals always the relation $p_r^{(eff)}=-\rho^{(eff)}$  for
both solutions. The tangential pressure has this behaviour only very near the origin of the radial co-ordinate. For both solutions, there is a small difference between these two pressures that grows as $r$ increases. This fact reveals the anisotropy of the effective matter content for the theory.
\section{Conclusion}\label{sec4}

In this paper, we have investigated the existence of regular black hole structures for a general $f(R)$ theory, sourced by
non-linear electromagnetic terms expressed by the Lagrangian $\mathcal{L}_{NED}$. Our approach follows very closely that one employed in Ref. \cite{zanchin}: instead of choosing specific forms for the
the $f(R)$ and $\mathcal{L}_{NED}$ functions, the approach consists in expressing the metric in terms of a mass function $M(r)$ and to choose a mass function that satisfies some requirements. Specifically, we have used the mass function 
determined in Ref. \cite{balart1}. Such mass function was constructed, in the context of GR theory coupled to non-linear electromagnetic field, in order to satisfy the WEC and to have an asymptotic Reissner-Nordstr\"om limit. In fact, the chosen mass function $M(r)$ is the most general functional form satisfying the WEC in GR.

Applied to the case of a general $f(R)$ and $\mathcal{L}_{NED}$ functions, that mass function of Ref. \cite{balart1} leads to regular black hole solutions which contain 
two horizons, the event horizon and the Cauchy horizon. We worked out completely two specific cases of that mass function, by choosing specific values for the free parameters in the model developed in Ref. \cite{balart1}. The regular character of the solutions is attested by the regular behaviour of the geometric invariants, like the Ricci scalar and the Kretschmann scalar.  Asymptotically, as expected,
the metric functions reproduces the Reissner-Nordstr\"om solution of the GR theory.

The energy conditions are satisfied for the two specific cases studied here, except for the case of the Strong Energy Condition (SEC) which is violated in
the vicinity of the Cauchy horizon. Of course, a violation of at least some of the energy conditions must occur if
regular solution must be extracted from the original theory. In this case, the violation is quite mild since it is only
the energy condition connected with the convergence of the geodesics that is violated (SEC), and even though in a quite
restricted region of the whole space-time.

Evidently, there are many open issues related to the problem treated here, like the complete determination of the $\mathcal{L}_{NED}$ function corresponding to the configurations found, and the stability problem. We postpone such new analysis to future works.

\vspace{1cm}

{\bf Acknowledgement}: MER 
thanks UFPA, Edital 04/2014 PROPESP, and CNPq,
Edital MCTI/CNPQ/Universal 14/2014,  for partial financial support. JCF thanks CNPq (Brazil) and FAPES (Brazil) for financial support.

\section*{Appendix: Asymptotic analysis of the new solutions}

Let us analyse the regularity of the solutions in the limit $r \rightarrow \infty$. The radial coordinate is redefined as $x=1/r$. Hence, $r\rightarrow\infty$ implies $x\rightarrow 0$. The metric function $e^{a(r)}$, for the solution \eqref{a2}, behaves in this limit as,
\begin{eqnarray}
e^{a(r)}\sim 1-2m x+q^2x^2+O(x^3).
\end{eqnarray}
From this behaviour, it is possible to verify that the metric behaves, up to second order, as in the Reissner-Nordstr\"om solution. 
In this limit we have,
\begin{eqnarray}
f(R)\sim \frac{1}{\pi^2q^2}\left[38 c_1\frac{m^3}{q^4}+O(x^5)\right].
\end{eqnarray}
Hence, the function $f(R)$ becomes asymptotically a constant, which depends on $c_1$, that can not be made zero.
The Lagrangian density $\mathcal{L}_{NED}$ becomes also a constant. We can verify this by using the expression \eqref{L} for the model \eqref{M1} which becomes,
\begin{eqnarray}
\mathcal{L}_{NED}\sim-\frac{192c_1m^3}{\pi^2 q^4\kappa^2}-\frac{c_1x}{\kappa^2}+O(x^3).
\end{eqnarray}
From this expression, it is possible to certify that the Lagrangian density becomes also a constant, which is multiplied, in the action, by $2\kappa^2$: it is zero in this limit, what is normal for solutions asymptotically flat. Performing the same analysis for the Lagrangian density given by \eqref{LF}, we find
\begin{eqnarray}
&&\mathcal{L}_F\sim -\frac{q^2\kappa^2}{c_1}x^3+O(x^4).
\end{eqnarray}

The component $F^{10}$ of the electric field (\ref{F10-2}) approximates to,
\begin{eqnarray}
&&F^{10}\sim \frac{3m}{\kappa^2 q}c_1-\frac{c_1}{\kappa^2qx}+\frac{2c_0q}{\kappa^2}x^2+O(x^3).
\end{eqnarray}
Remark that there is a divergence. But this divergence does not affect the physical quantities neither the equations of motion for this solution. In order to verify this, we will evaluate the electric energy density.
Hence, let us turn the attention to other physical quantities like the electrical energy density (right side to Eq. (\ref{eq1})), which becomes,
\begin{eqnarray}
\rho \sim \frac{192m^3}{\pi^2 q^2}c_1+2c_1x+O(x^2). 
\end{eqnarray}
The electrical energy density goes to a constant in the limit $r\rightarrow+\infty$. On the other hand, the effective electric energy density is given, in this limit, by
\begin{eqnarray}
\rho_{eff} \sim \frac{q^2}{\kappa^2}x^4+O(x^5), 
\end{eqnarray}
which goes asymptotically to zero.

For the solution \eqref{a2} the left side for first equation of motion, in the limit $r\rightarrow +\infty$, leads to,
\begin{eqnarray}
192c_1\frac{m^3}{\pi^2 q^4}+2c_1x-3c_1mx^2+O(x^3).
\end{eqnarray}
This result shows that the infinite quantity coming from $f_R(r)=c_0+c_1r$ does not affect the equations of motion. The same happens for the second equation of motion. For the third equation of motion, we find
\begin{eqnarray}
192c_1\frac{m^3}{\pi^2 q^4}+c_1x+O(x^3).
\end{eqnarray}
Hence, the equations of motion are regular in the limit  $r\rightarrow+\infty$: the equations are consistent since the infinity coming from $f_R(r)=c_0+c_1r$ is compensated by other terms.
\\
Now, we perform a similar asymptotical analysis ($r\rightarrow+\infty$, with $x=1/r$) for the second solution given by (\ref{a3}). In this case,
\begin{eqnarray}
e^{a(r)}\sim 1-2mx+q^2x^2+O(x^3),
\end{eqnarray}
showing the asymptotically the metric coincides with the Reissner-Nordstr\"om. 

The function $f(R)$ for this solution reads, 
\begin{eqnarray}
f(R)\sim 24c_1\frac{m^3}{q^4}+O(x^5).
\end{eqnarray}
Again, the function $f(R)$ becomes asymptotically a constant which depends on $c_1$, which can not be made zero. 
Inspecting the action, we have that $f(R)\sim cte$ or $f(R)\sim -2\Lambda$. On the other hand, the component $F^{10}$ becomes,
\begin{eqnarray}
F^{10}\sim -\frac{c_1}{q\kappa^2x}+\frac{3c_1m}{q\kappa^2}+\frac{2c_0q}{\kappa^2}x^2+O(x^3),
\end{eqnarray}
Hence, the component $F^{10}$ diverges in the limit $r\rightarrow+\infty$. But this does not imply the presence of divergences in quantities like the energy density, among others, neither in the equations of motion. 
In fact, the Lagrangian $\mathcal{L}_{NED}$ becomes
\begin{eqnarray}
\mathcal{L}_{NED}\sim -12c_1\frac{m^3}{q^4\kappa^2}-\frac{c_1}{\kappa^2}x+O(x^3),
\end{eqnarray}
leading to,
\begin{eqnarray}
\mathcal{L}_F\sim -\frac{q^2\kappa^2}{c_1}+O(x^4).
\end{eqnarray}
For the electrical energy density we find
\begin{eqnarray}
\rho\sim 12c_1\frac{m^3}{q^4}+2c_1x-3c_1mx^2+O(x^3),
\end{eqnarray}
which is a constant in the limit $r\rightarrow+\infty$.
For the effective energy density, we find,
\begin{eqnarray}
\rho_{eff}\sim \frac{3q^6}{4m^2\kappa^2}x^6+O(x^7),
\end{eqnarray}
which goes to zero in the limit $r\rightarrow+\infty$. 

Let us now verify how the equations of motion behave. The left hand side of the first one reads in that limit, 
\begin{eqnarray}
12c_1\frac{m^3}{q^4}+2c_1x-3c_1mx^2+O(x^3).
\end{eqnarray}
This shows that the infinity introduced by 
$f_R(r)=c_0+c_1r=c_0+c_1x^{(-1)}$ is cancelled, and does not contribute to the equations of motion. The same happens for the second equation (by symmetry). For the third equation, we find,
\begin{eqnarray}
12c_1\frac{m^3}{q^4}+c_1x+O(x^3).
\end{eqnarray}
Hence, we confirm that a regular behaviour is also verified for the second solution, since there is no divergence in the physical relevant quantities neither in the equations of motion.


\end{document}